\documentclass[
    ,final            % use final for the camera ready runs
  ]{aipproc}

\layoutstyle{6x9}

\usepackage{amssymb}
\usepackage{epstopdf}
\usepackage{color}
\def\mjy{ {\rm mJy}}
\def\lsi{ {LS~I~+61~303}}

\begin{document}

\title{First LOFAR Observations of Gamma-Ray Binaries}
%%%%% Keywords??: ask to Marc and Josep.
\classification{98.70.Rz; 98.70.Dk; 98.70.Qy}%<Replace this text with PACS numbers; choose from this list:
               %\texttt{http://www.aip..org/pacs/index.html}>}
\keywords      {gamma rays: binaries -- radio continuum:stars -- radiation mechanism: non-thermal}

\author{B.~Marcote}{
  address={Departament d'Astronomia i Meteorologia, Institut de Ci\`{e}ncies del Cosmos (ICC),
  Universitat de Barcelona (IEEC-UB)}
}
\author{M.~Rib\'o}{
  address={Departament d'Astronomia i Meteorologia, Institut de Ci\`{e}ncies del Cosmos (ICC),
  Universitat de Barcelona (IEEC-UB)}
}
\author{J.~M.~Paredes}{
  address={Departament d'Astronomia i Meteorologia, Institut de Ci\`{e}ncies del Cosmos (ICC),
  Universitat de Barcelona (IEEC-UB)}
}
\author{J.~Swinbank}{
  address={Astronomical Institute ``Anton Pannekoek'', University of Amsterdam}
}       
\author{J.~Broderick}{
  address={School of Physics and Astronomy, University of Southampton}
}
\author{R.~Fender}{
  address={School of Physics and Astronomy, University of Southampton}
}
\author{S.~Markoff}{
  address={Astronomical Institute ``Anton Pannekoek'', University of Amsterdam}
}       
\author{R.~Wijers}{
  address={Astronomical Institute ``Anton Pannekoek'', University of Amsterdam}
}

\begin{abstract}
 A few binary systems display High Energy ($100\ {\rm MeV}-100\ {\rm GeV}$) and/or Very High Energy ($\gtrsim 100\ {\rm GeV}$) gamma-ray emission. These systems also display non-thermal radio emission that can be resolved with long-baseline radio interferometers, revealing the presence of outflows.
   It is expected that at very low frequencies the synchrotron radio emission covers larger angular scales than has
     been reported up to now.
       Here we present preliminary results of the first deep radio observations of the gamma-ray binary \lsi\ with LOFAR, which is sensitive to extended structures on arcsecond to arcminute scales.
\end{abstract}

\maketitle

\section{Introduction}

Less than 10 binary systems have been reported as High Energy (HE: $100\ {\rm MeV}-100\ {\rm GeV}$) and/or Very High Energy (VHE: $\gtrsim 100\ {\rm GeV}$) gamma-ray emitters. All of them consist of a young massive star and a compact object (either neutron star or black hole). Some of these systems are X-ray binaries that accrete matter from their companion stars, and display a non-thermal spectral energy distribution (SED) dominated by the X-ray photons. The two known sources belonging to this class are Cygnus X-3, clearly detected by both {\em Fermi} \citep{fermi2009} and {\em AGILE} \cite{tavani2009}, and Cygnus X-1, possibly detected by {\em AGILE} \cite{sabatini2010} and MAGIC \cite{albert2007}.

There is another family of these systems in which the SED is dominated by the 
% Gamma-ray binaries are binary systems consisting of a compact object and a young massive star which have a non-thermal spectral energy distribution (SED) dominated by
MeV-GeV photons (see \citep{paredes2011} and \citep{dubus2006}). In analogy with the definition of X-ray binary, we refer to these sources as gamma-ray binaries.
Their emission is displayed across all the electromagnetic spectrum, from radio to VHE gamma rays. Up to now, five Galactic gamma-ray binary systems are known: PSR~B1259$-$63, LS~5039, LS~I~+61~303, HESS~J0632+057 and 1FGL~J1018.6$-$5856.
Only in the case of PSR~B1259$-$63 the nature of the compact object is known: a young non-accreting pulsar displaying a strong relativistic wind \cite{johnston1992}. In the other cases, the nature of the compact object, either black hole or neutron star, is unknown, but up to now there is no evidence of the presence of an accretion disk in these binaries.
%The physical processes around them, either accretion or 
%The physical properties of these powerful accelerators are still under discussion.
In this observational context, two competing scenarios have been proposed to explain the broadband emission from gamma-ray binaries: the {\em microquasar scenario} and the {\em binary pulsar scenario}. %shock between the wind of the pulsar and the stellar wind}.
The first one involves an accreting black hole or neutron star with relativistic jets (displayed by many X-ray binaries), and the second one a young non-accreting pulsar with a shock between its wind and the wind (and/or circumstellar envelope) of the massive star.

In both, X-ray and gamma-ray binaries, we detect synchrotron radio emission that has been resolved by interferometers at centimeter wavelengths.
%Both scenarios can explain the synchrotron radio emission detected and resolved by interferometers at centimeter wavelengths.
For example, Cygnus~X-3 has been resolved with the VLA \cite{marti2008} and the VLBA \cite{miller-jones2004}, while PSR~B1259$-$63 has been resolved with LBA \cite{moldon2011}.
In the gamma-ray binaries LS~5039 and LS~I~+61~303 extended radio emission at mas scales showing periodic morphological variability has been observed \cite{moldon2012}\cite{dhawan2006}\cite{massi2012}\cite{javi2012thesis}.

%In the binary pulsar scenario,
At meter wavelengths we should observe the synchrotron radio emission from the low-energy electrons. %%%%%%%%%%%%%%%%%%%%%%%%%
%which should remain unaffected by energy looses on any plausible dynamical timescale \cite{durant}.
In the binary pulsar scenario this emission would be located in a region farther away from the binary system than at higher frequencies. These electrons should 
%These electrons should belong to larger structures on the system, that
remain unaffected either by inverse compton or synchrotron cooling \cite{durant2011}. Therefore, we would expect to see synchrotron radio emission on much larger angular scales than it has been reported up to now, without variability along the orbit \citep{valenti2011}.
    This emission has remained unobserved due to the very poor resolution of the previous radio observatories at these very low frequencies.
    % Revisar: escenarioo Durant solo para pulsar, no para microquasars.

\section{LS~I~+61~303}

\lsi\  is a gamma-ray binary discovered as a gamma-ray emitter 30 years ago, with periodic radio outbursts coincident with its orbital period of $P = 26.5\ {\rm days}$ \citep{gregory2002}.
Although both the microquasar and the binary pulsar scenario have been proposed for this source, the VLBA observations showing periodic morphological changes \cite{dhawan2006} are typically interpreted as evidence of the second scenario.
In this case we could expect to detect extended radio emission at very low frequencies \cite{durant2011}.
%However, the origin of its radio emission is still unclear, and it could be due to a jet or to an outflow produced after the shock between the winds of the putative pulsar and the star. This last scenario has been proposed to explain the VLBA observations \citep{dhawan}, where periodic morphology changes, showing a cometary tail, have been reported.
We note that evidence of an extended X-ray structure has been reported in this source \citep{paredes2007}.

From radio observations, \lsi\ displays periodic orbital variability in the range of $1$ to $10\ {\rm GHz}$ \cite{gregory2002}\cite{strickman1998}. At lower frequencies, from $260$ to $\sim 700$ MHz, variability on $\sim 1$-year scales has been reported \cite{pandey2007}.
%Radio observations with the VLA have been conducted through a complete orbital period \citep{strickman}, showing a periodic flux variability. This variability has been detected from $\sim {\rm GHz}$ frequencies to $\sim 300\ {\rm MHz}$. Observations from GMRT at 235 and 610~MHz have also been conducted \citep{pandey}. Variability at these frequencies, is also reported.
If this variability at the lower frequencies follows the orbital period or has larger timescales is unknown. 
%As mentioned above, at the very low frequencies is expected to find a persisting emission at larger scales, without variability.
%If this behaviour persists going down to lower frequencies is unknown at this state of the art. A persisting emission at larger scale, without variability, is expected to merge at the very low frequencies.

\section{LOFAR}

  The Low Frequency Array (LOFAR) is a digital radio interferometer with stations in The Netherlands, France, Germany, Sweden and the United Kingdom which will
  %observe continuously a large fraction of the northern hemisphere 
  regularly monitor a large fraction of the northern sky
  (see \citep{heald2011} for a recent update). LOFAR detects photons in the frequency range 30--240 MHz, which has never been explored by any large-scale interferometer before. Operating in this new frequency window LOFAR promises to revolutionize wide ranging areas of astrophysics.
    LOFAR has some Key Science Projects (KSPs) defined to drive the design of its operational performance. These KSPs are:
%    \begin{itemize}
%        \renewcommand{\labelitemi}{$\circ$}
%        \vspace{-3pt}
Epoch of Reionisation, Deep Extragalactic Surveys, Transient Sources and Pulsars, Ultra High Energy Cosmic Rays, Solar Science and Space Weather, and Cosmic Magnetism.
%        \vspace{-3pt}
%   \end{itemize}
  The LOFAR radio telescope consists of many low-cost antennas, distributed in 24 core, 18 remote and 8 international stations, with baselines from 100~m to 1\,500~km. There are two types of antennas: Low Band Antennas (LBA) observing at 30--80 MHz and the High Band Antennas (HBA), in the range 120--240 MHz.
  LOFAR started preliminary observations in 2008, and detailed commissioning observations in 2011. In the final configuration, % When all the international stations will be working,
the resolution and sensitivity of LOFAR will reach $0.7\ {\rm arcsec}$ and $10\ \mjy\ {\rm beam}^{-1}$ at $60\ {\rm MHz}$ or $0.2\ {\rm arcsec}$ and $0.3\ \mjy\ {\rm beam}^{-1}$ at $240\ {\rm MHz}$ in one hour of observation.

\section{Observations and Results}

We present the first LOFAR observation of LS~I~+61~303, made on September 30, 2011 for a 6-hour run with the HBA at 120--180 MHz. For this observation 23 core stations + 9 remote stations were used. The orbital phase of \lsi\ during the observation was 0.55 (the periastron takes place at 0.23 \cite{casares2005}).
  The data were analyzed with the LOFAR Standard Imaging Pipeline and CASA software (NRAO). We have used standard procedures developed by the LOFAR team during the commissioning stage for the calibration, and the CASA routines for the imaging.
  
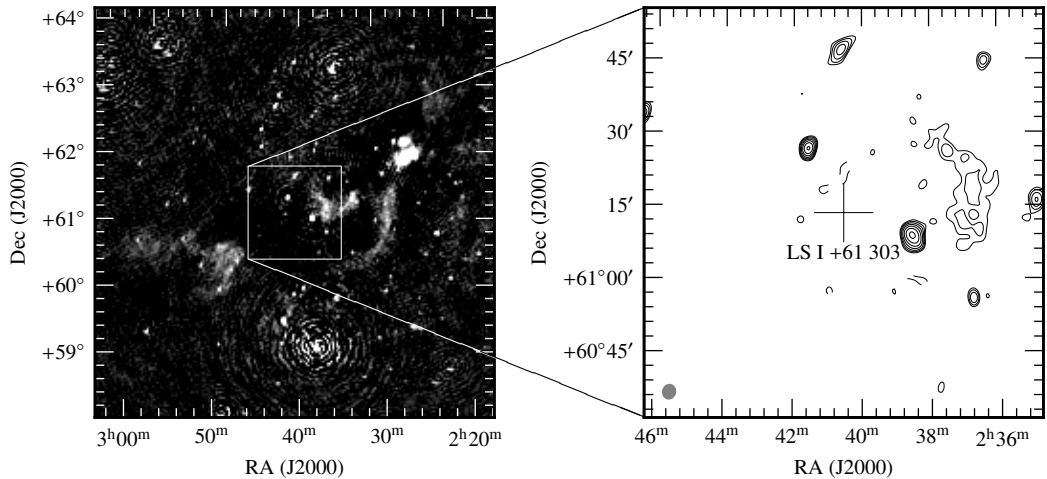
\begin{figure}
    %\includegraphics[height=0.31\textheight,trim=27 29 27 26]{images/fig-contours.eps}
    %\includegraphics[height=0.31\textheight,trim=10 10 10 10]{images/fig-contourDetailWithMarker2.eps}
    %\put(-312,83){\line(2,-1){118}}
    %\put(-312,125){\line(2,1){118}}
    %\put(-359,83){\framebox(47,42){ }}%\vspace{-2pt}
    \includegraphics{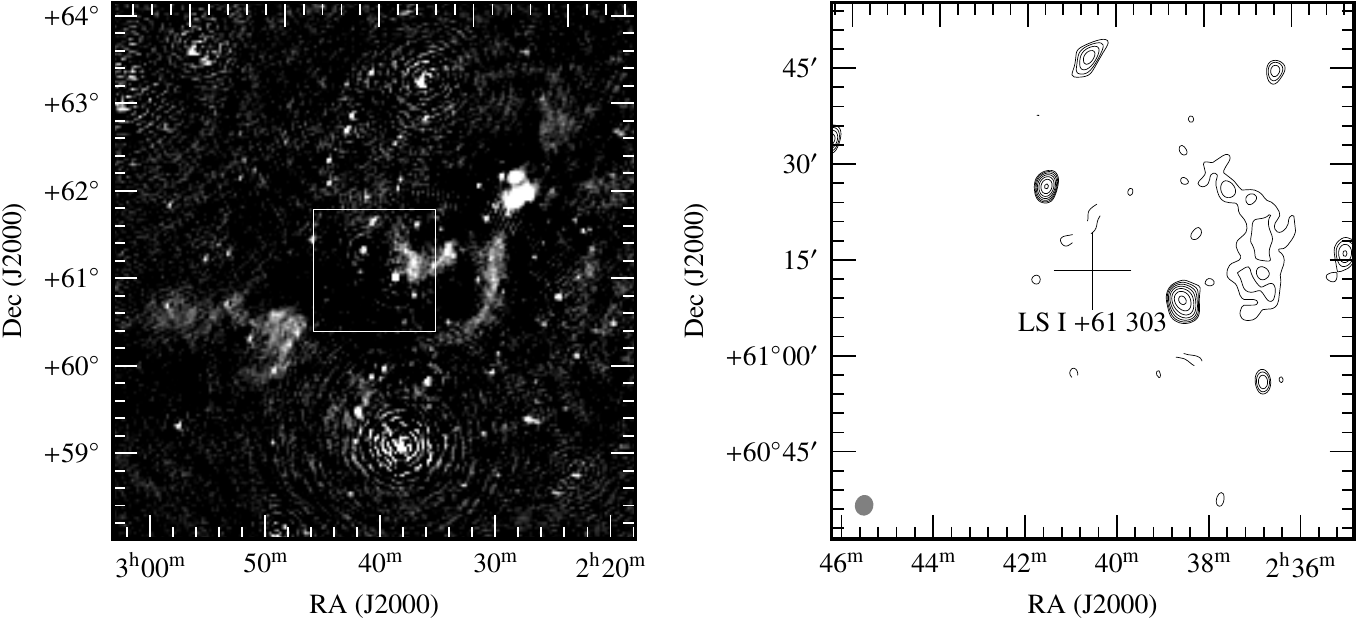}
    \put(-301,118.8){\color{white}\line(5,2){95}} %118
    \put(-301,83.5){\color{white}\line(5,-2){95}}
    \put(-213.5,154){\color{black}\line(5,2){62}} %118
    \put(-213.5,48.5){\color{black}\line(5,-2){63}}
    %\put(-359,83){\framebox(47,42){ }}%\vspace{-2pt}
    % CAPTION AND PUT LINES FROM ONE IMAGE TO THE OTHER (THE ZOOM)
    \caption{Preliminary image of \lsi\ from a 6-hour observation with LOFAR at 120--130 MHz. On the {\bf left} we can see a large fraction of the primary beam ($6\ {\rm degrees}$) and on the {\bf right} we show a $1.4\times1.4$ degrees zoom around \lsi. The synthesized beam has a size of $\sim 2.5\ {\rm arcmin}$ and is shown in the bottom-left corner of the right image. The $rms$ of the image is $ \approx 40\ \mjy\ {\rm beam^{-1}}$ around the \lsi\ position and the contours start at $3\sigma$ $rms$ level.\vspace{-20pt}}
    \label{fig-obs} 
\end{figure}

  Figure~\ref{fig-obs} shows the resulting image for the integration of the frequency range 120--130 MHz, with $uv$-distances shorter than 100$\lambda$ removed to avoid the introduction of extended emission from the Galactic plane. The primary beam is $\sim 6\ {\rm degrees}$.
 There is no significant excess at the position of \lsi, with a $rms \approx 40\ \mjy\ {\rm beam}^{-1}$ and a resolution of $\approx 2.5~{\rm~arcmin}$.
Extrapolating the results of \citep{strickman1998}, we expect a flux density of $\approx 44\ {\rm mJy}$ at $125\ {\rm MHz}$. This is close to the $rms$ of our preliminary image. However, in the final image including all subbands we expect to have a $rms$ of $\sim 10\ \mjy\ {\rm beam}^{-1}$, and therefore \lsi\  should be detectable above a $3$-$\sigma$ confidence level.
%According to some theoretical estimations derived from \citep{valentiLS}, we could expect emission at scales from $\sim$ arcsec up to $\sim$ arcmin at these frequencies. At higher frequencies this extended structure dissapears. It is not detected by VLA observations at 5~GHz \citep{marti1998} % conducted deep VLA observations at 5 GHz, where no extended structure is detected
%in arcsecond to arcminute scales.

Any extended structure in arcsecond to arcminute scales is not detected in deep VLA observations at 5 GHz with a $rms \gtrsim 0.1\ {\rm mJy\ beam^{-1}}$ \cite{marti1998}. However, according to the theoretical models discussed above \cite{durant2011,valenti2011}, we could expect emission at scales from ~arcsec up to ~arcmin at very low frequencies.

Although a better performance is required to detect \lsi, LOFAR will experience significant improvements during the following years, with the addition of international stations and future software developments.
At the end of the commisioning stage, LOFAR will have a much better sensitivity and angular resolution for low-frequency observations.
This will allow us to determine if the variability along the orbit seen at higher frequencies in \lsi\  is still present or not at low frequencies and if extended emission from gamma-ray binaries is plausible.

\begin{theacknowledgments}

We thank the staff of LOFAR who have made these observations possible. LOFAR, designed and constructed by ASTRON, has facilities in several countries, that are owned by various parties, and are collectively operated by the International LOFAR Telescope (ILT) foundation under a joint scientific policy. The CASA software is developed by the National Radio Astronomy Observatory (NRAO).
  B.M., M.R. and J.M.P. acknowledge support by the Spanish Ministerio de Econom\'{i}a y Competitividad (MINECO) under grants AYA2010-21782-C03-01 and FPA2010-22056-C06-02. B.M. acknowledges financial support from MINECO under grant BES-2011-049886. M.R. acknowledges financial support from MINECO and European Social Funds through a Ram\'{o}n y Cajal fellowship. J.M.P. acknowledges financial support from ICREA Academia.
%% Extended version
%We thank the staff of LOFAR who have made these observations possible. 
%LOFAR, the Low Frequency Array designed and constructed by ASTRON, has facilities in several countries, that are owned by various parties (each with their own funding sources), and that are collectively operated by the International LOFAR Telescope (ILT) foundation under a joint scientific policy. The CASA software is developed by the National Radio Astronomy Observatory (NRAO).
% B.M., M.R. and J.M.P. acknowledge support by the Spanish Ministerio de Econom\'{i}a y Competitividad (MINECO) under grants AYA2010-21782-C03-01 and FPA2010-22056-C06-02. B.M. acknowledges financial support from MINECO under grant BES-2011-049886. M.R. acknowledges financial support from MINECO and European Social Funds through a Ram\'{o}n y Cajal fellowship. J.M.P. acknowledges financial support from ICREA Academia.
\end{theacknowledgments}

\bibliographystyle{aipproc}   % if natbib is available
%\bibliography{/home/gamma/folre/bmarcote/Documents/Mendeley/library.bib}

\end{document}